\documentclass{elsart}
\usepackage{epsfig}
\newcommand{\g}{$\gamma$}

\newcommand{\bc}{\begin{center}}
\newcommand{\ec}{\end{center}}
\newcommand{\be}{\begin{equation}}
\newcommand{\ee}{\end{equation}}
\newcommand{\bfg}{\begin{figure}}
\newcommand{\efg}{\end{figure}}
\newcommand{\bi}{\begin{itemize}}
\newcommand{\ei}{\end{itemize}}
\newcommand{\bt}{\begin{table}}
\newcommand{\enta}{\end{table}}

\renewcommand{\deg}{\ensuremath{^\circ}}

\begin{document}
\begin{frontmatter}
\title{Can a simple instrument measure polarization in gamma-ray bursts?} 
\author{A.~Curioni},
\ead{curioni@astro.columbia.edu}
\author{N.~Mirabal}
\ead{abulafia@astro.columbia.edu}
\address{Columbia University, New York, NY, USA}

\begin{abstract}
The recent claim of a high degree of linear polarization in the prompt gamma-ray
burst (GRB) emission of GRB 021206 might have important implications for the
underlying mechanism ultimately responsible for the GRB radiation. 
While the claim itself remains controversial, a full characterization of the  
GRB polarization has become a scientific imperative. A review of past and
present polarimetry missions motivates a set of guidelines for future dedicated GRB
polarization experiments. It is also argued that polarization in GRBs could be
measured by a relatively simple instrument using readily available technology.  
\end{abstract}

\begin{keyword}
gamma-ray \sep polarization

\PACS 
\end{keyword}
\end{frontmatter}

\section{Introduction}

The nature of the prompt GRB emission remains one of the outstanding
astrophysical mysteries of the past 35 years \cite{kle}. The central problem
concerns the lack of a proven mechanism that can both extract energy from a
collapsing GRB progenitor, and generate prime conditions for the production of 
relativistic outflows \cite{meszaros}. 
Three models have emerged as the leading candidates for explanation of the
prompt GRB emission. In conventional hydrodynamic models, internal or external
shock fronts accelerate particles that radiate high-energy photons
\cite{rees,meszaros-93,dermer,piran}.  
In the case of electromagnetic models, the GRB progenitor loses much of its spin
in the form of an  electromagnetically dominated outflow that can extend all the
way out to the \g-ray emission region \cite{lyutikov}. The cannonball model relies
on inverse Compton scattering of photons to GRB energies \cite{dar}. 
Although various numerical simulations have been  successful in reproducing 
the complex structure observed in GRB light curves \cite{koba1997,liang2003},
the observations needed to validate any of these models are still sparse.  

Perhaps the greatest observational challenge for prompt emission models has been
put forward by the recent claim of polarization at the $\Pi$ = 80 $\pm$ 20\%
level in observations of GRB 021206 by $RHESSI$ (Ref.~\cite{coburn2003}, CB03 
hereafter).  
While the interpretation of the data remains controversial (Ref.~\cite{rut2003},
RF03 hereafter, and Ref.~\cite{boggs03}, BC03), it is clear that a significant
measurement of polarization in GRBs might provide fundamental insight into the
GRB mechanism. 
For example, electromagnetic models could be favored if $\Pi \approx$ 30--40\%
is established \cite{lyutikov,granot}, since hydrodynamical models have
difficulties generating more than 10\% fractional polarization (but see
Ref.~\cite{nakar}). Only the cannonball model with some fine tuning may
accommodate fractional polarization as large as $\Pi \approx$ 80\%.  
Any measurement of $\Pi <$ 20\% would not exclude any of the three models
since for each of them depolarization effects can enter into play.  
Nonetheless, an upper limit in polarization would help placing stringent
boundaries for the ongoing theoretical effort.  

The exciting prospects of polarization measurements in the prompt GRB emission
is moving research into new directions \cite{mcconnell2}. In particular,
consensus is growing that a dedicated  polarization experiment is an inevitable
step in constraining the GRB mechanism. 
In this work, after a general introduction (Sec.~2), we discuss the general
guidelines for any dedicated GRB polarization experiments in the future
(Sec.~3); we also argue that the technology is readily available to develop a
new generation of instruments that can achieve improved sensitivity and sky
coverage (Sec.~4); lastly, we outline the basic design for a relatively simple
experiment that meets the conditions for a significant polarization measurement
(Sec.~5).   

\section{Compton Scattering and Polarimetry}

Polarization measurements in the soft \g-ray band, 0.2-2~MeV, commonly rely on
Compton polarimetry \cite{mcconnell}. Although practical applications may be
difficult to implement, the underlying physics is already in the differential 
cross-section for Compton scattering as given by the Klein-Nishina formula

\begin{equation}
\frac{d\sigma}{d\Omega} = \frac{r_0 ^2}{2} \left( \frac{E'}{E_0} \right)^2
\left( \frac{E'}{E_0} + \frac{E_0}{E'} - 2 \sin ^2 \theta \cos ^2 \eta \right) 
\end{equation}

where $r_0$ is the classical radius of the electron, $E_0$ the energy of the
incident photon, $E'$ the energy of the scattered photon, $\theta$ the scatter
angle and $\eta$ is the azimuthal angle between the direction of the {\bf E}
vector of the incident photon (i.e. its polarization) and the plane defined by 
the direction of the incident photon and the direction of the scattered
photon (Fig.~1).
Unpolarized radiation will show no dependence on $\eta$ once the effect has
been averaged out over many incident photons. 
Polarized photons, on the other hand, have a well-defined direction of the {\bf
E} vector, which implies that the scattering probability has a maximum for
$\eta$=90\deg and minima at $\eta$=0\deg and $\eta$=180\deg. Thus, photons are
more likely to be scattered in the plane normal to the  polarization vector. 

\begin{figure}[!ht]
\centering
\includegraphics[bb=150 490 500 725,width=0.8\textwidth,clip]{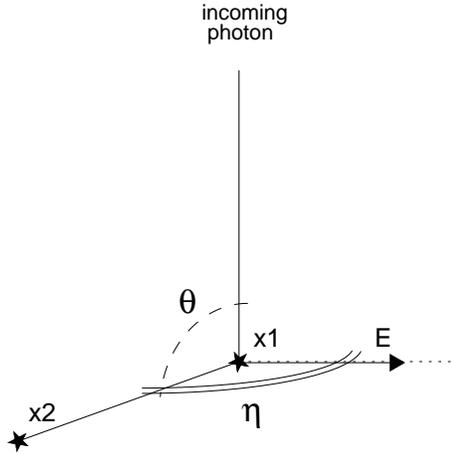}  
\caption{\label{fig-principle} Sketch of the geometry of Compton scattering,
including polarization. }
\end{figure}

A Compton polarimeter is any instrument able to measure the $\eta$-modulation of
Compton scattered photons; simplifying to the extreme, it only needs to record
the position of the first and second interaction for each Compton scattered 
photon \footnote{En passant, the reader should notice that these requirements
are only a subset of the requirements for Compton imaging, where also the energy
and the correct time sequence must be determined.}.
Because of the $\sin ^2 \theta$ factor in Eq.~1 the effect of polarization
is maximum for Compton scatter angles $\theta \approx$ 90\deg, therefore the
sensitivity of any reasonable Compton polarimeter should be optimized for
detecting events with such scatter angles. 
The dataset is given by a sample of positions ({\bf x$_1$}, {\bf x$_2$}) of the
first and second interaction for Compton scattered \g-rays, which we define as
double interactions. If now we look at the angular distribution of the vectors
{\bf x$_1$}-{\bf x$_2$} in the plane normal to the direction of the incoming
photons, given a fractional polarization $\Pi$, the signal $S$ as a function of
the angle $\eta$ will show a sinusoidal modulation     

\begin{equation}
\frac{dS}{d\eta} = \frac{S}{2 \pi} \left[ 1 - \mu \Pi \cos (2(\eta -
\phi)) \right]
\end{equation}

where $\mu$ is the instrumental modulation factor and $\phi$ is the polarization
angle of the incident photons \cite{mcconnell,novick1975}. 

Now assuming that $B$ is the total number of background events (un-modulated
with respect to $\eta$), the expected signal-to-noise ratio for polarization
measurements $\sigma$ can be written as   

\begin{equation}
\sigma = \frac{\mu \Pi S}{\sqrt{2(S+B)}}
\end{equation}

Therefore, the minimum detectable fractional polarization $\Pi _{min}$ (at the
$n _{\sigma}$ significance level) can be shown to be

\begin{equation}
\Pi _{min} = n _{\sigma} \frac{\sqrt{2(S+B)}}{\mu S}
\end{equation}

where the position of the source in the sky is tacitly included in the
instrumental modulation factor $\mu$. 

\section{Considerations for Future GRB Polarization Experiments}

Our previous discussion indicates that the design requirements for future GRB
polarization experiments can be summarized as: sensitivity in a suitable energy 
range, minimized background $B$, maximized signal $S$, and maximized $\mu$. 

Let us start with the energy range. While Compton scattering can occur at lower
energies and the Compton cross-section actually has a broad maximum at about
100~keV, it is - for most materials - the dominant process in some energy band
centered around $\sim$1~MeV. For example, for carbon, $Z$=6, Compton scattering
is the main contribution to the total cross section between 100~keV and 10~MeV. 
This energy range is well-suited for GRBs since the peak energy of the GRB
spectrum E$_{peak}$ is narrowly distributed around E$_{peak}$ $\approx$
250~keV \cite{fishman2001}. 

The requirement of a low background level is actually more relaxed than
in traditional MeV \g-ray observations (e.g. \cite{AJDean:91}). The fact that
the prompt GRB emission needs to be integrated over a relative short interval of
time, corresponding to the duration of the burst, significantly reduces the
relevance of the atmospheric \g-ray background and of the background due to
neutron interactions, spallation and activation.
Both CB03 and RF03 show that the most dangerous source of background is the
chance coincidence of two independent events, each of them giving only one
interaction, within the same time window $\tau$ that defines an event. Such an
occurrence clearly fakes a genuine double event. 
Fortunately, the dependence of the rate of chance coincidence events $r_B$ on
the source rate $r_{source}$ (detected rate of single events) and $\tau$ may be
written as  
\begin{equation}
r_B = r_{source} ^2 \times \tau 
\end{equation}
Thus $r_B$ may be made negligible provided a sufficiently narrow time window
$\tau$. For example, the COMPTEL instrument exploited a time-of-flight
measurement to reconstruct the direction of the \g-rays, with a timing accuracy
better than $\sim$1 ns \cite{scho}.  
Conservatively assuming that each photon is individually time tagged with a
accuracy of $\tau \sim$10 ns, a source rate $r_{source}$ as high as 10~kHz
implies a negligible rate of chance coincidences $r_B \approx$ 1 Hz.
The fast timing requirement suggests that {\it scintillator counters} might be a
logical choice for future GRB polarization experiments.  

The signal $S$ is defined as 
$$
S = F \times A _{eff} \times T
$$
where $F$ is the flux impinging of the detector in units of photons cm$^{-2}$
s$^{-1}$, $A _{eff}$ is the effective area and $T$ is the duration of the GRB. 
Since $F$ and $T$ are intrinsic properties of the GRB, a large $S$ can be
accomplished only by increasing $A _{eff}$. 
The NaI array of BATSE Large Area Detectors (LADs) allowed for the entire sky to 
be viewed simultaneously with an effective area $A _{eff} \approx$ 2000 cm$^2$
\cite{fishman89}. Taking BATSE as the prototypical sensitive detector for GRBs,
we assume that a comparable effective area must be achieved. It should be
noticed that the effective area for double interactions is only a fraction of
the effective area {\it \`a la} BATSE. In fact, a \g-ray not only has to
interact in the detector (through Compton scattering), but also the scattered
\g-ray must re-interact and be separately detected. More details on how to
achieve a large effective area are given in Sec.~4.

Finally, the modulation factor $\mu$ should be no worse than the 
$\mu \approx$ 20\% already achieved by $RHESSI$ (CB03), which suggests that a
similar geometrical arrangement with a fixed number of cylindric detectors
distributed over a planar support-structure is already advantageous.  

Before proceeding further in the discussion, it is worthwhile stressing that,
following approach in CB03 and RF03 and their analysis of the $RHESSI$ data,
there are no specific requirements on energy resolution and source imaging
capability of the polarimeter itself.

\section{Basic Design of an Advanced GRB Polarimeter}

Before discussing the proposed design in detail, it is convenient to review some
characteristics of $RHESSI$.
Although primarily designed for the study of the physics of particle 
acceleration in solar flares, $RHESSI$ has proven sensitive to GRB polarization
measurements using its germanium focal plane detectors. Oddly enough, of the
approximately 80 GRBs detected by RHESSI, detailed analysis has only yielded
one seemingly plausible polarization measurement in the case of GRB 021206
(CB03).  
It is possible that the report of polarization in GRB 021206 has been
overestimated (RF03); however, the detection of polarization 
only in the case of GRB 021206 is not entirely surprising given its unusual 
properties, i.e. a $<$ 1\deg~localization, a fluence ranking in the top 5\%
among all observed GRBs \cite{hurley1}, and the fact of being only slightly
off-axis with respect to $RHESSI$.

Upon careful examination, $RHESSI$ offers several advantages over alternative
polarimetry configurations.  
We are thus lead to consider a detector design that combines the geometrical
arrangement of $RHESSI$, and an effective area comparable to BATSE as the basis
for a future GRB polarimetry experiment. \\
As the ``building block'' of such an instrument we consider a cylindric detector
unit made of scintillator material coupled with possibly one single
photodetector (e.g. a standard photomultiplier tube). This would play the same
role as the $RHESSI$ Ge detectors. Like in the CB03 analysis, the interaction
location corresponds to the position of the detector unit. In $RHESSI$, the
distance between the centers of two neighboring detector units is about twice
the diameter of the detector unit itself, and we assume the same geometry.  

The main constraints on the design of the cylindric detector unit derive from
the requirement of a large effective area for double interactions. Here a
configuration as in Fig.~2 is assumed, and $r$ is the radius of the cylinder and 
$t$ its height. 
A large geometrical area $\pi r^2$ and enough ``thickness'' $t$ (e.g. several 
mass attenuation lengths \footnote{The mass attenuation length is defined as
the intensity $I$ remaining after traversal of a thickness $t$ (in units of
mass/unit-area) of the specified material: $I=I_0 e^{-t/\lambda}$. It depends
on the energy of the \g-ray. A large database is available at
http://physics.nist.gov/PhysRefData. } so that $>$90\% of the impinging flux 
will interact within the detector) are the starting point for a 
large effective area. 
Since we are interested in double interactions, we now want to maximize the
number of double events interacting in two separate detector. For this reason, 
the radius of a detector unit should be about 1 mass attenuation length, in
order to guarantee that 
\begin{itemize}
\item[1.] a large fraction of the Compton scattered photons does reach a second
detector without being absorbed within the same detector unit; 
\item[2.] once a scattered photon reaches a second detector unit, it has a
large enough probability to interact. 
\end{itemize}

Low $Z$ materials would be well suited for building a large area detector unit
minimizing the probability of a double interaction within the same unit. 
Organic scintillators (plastic or liquid, possibly loaded with high $Z$
elements) are usually compounds or mixtures of low $Z$ elements and they would
fully meet the requirement of fast timing capability. They are also
extremely cost-effective, easy to operate and easily machinable for building
large area detectors, as opposed to costly Ge detectors. Their fast (ns)
response makes them ideal for applications which require an excellent timing
capability. \\  
The one drawback in using low $Z$ materials in general and organic scintillators
in particular is that the mass attenuation length would be too large (10 cm or
more for non-loaded organic scintillators at \g-ray energies around the
Compton minimum in the cross-section, i.e. $\sim$1~MeV), which might make the
size of the detector impractically large.  
How serious this problem is needs to be studied in greater detail, considering
organic scintillators as NE 226 which has a relatively high density (1.61) and
is well suited for detecting \g-rays, or NE 316, Sn loaded. 

 
Given a $RHESSI$-like geometrical arrangement with $n$ counters, we define the
total effective area as  

\begin{equation}
A _{eff} = n \pi r^2 ~( 1 - e^{-t/\lambda} )
\end{equation}

i.e. we consider an on-axis GRB. The mass attenuation length $\lambda$ should be
evaluated at a few hundreds keV, given the spectral properties of most GRBs, and 
we assume $r \approx  \lambda$. To make $(1 - e^{-t/\lambda}) \approx 1$, the
thickness has to be $z/\lambda ~\ge$~3. A large thickness also increases the
solid angle for scattered photons to hit a second detector, therefore improving
the efficiency to Compton scattered events.  
For plastic or liquid scintillator counters, we may assume $r \sim$7 cm, and as
few as $n=$15 would already give a 2000 cm$^2$ effective area {\it \`a la}
BATSE. It is easy to imagine a scaled up version, with an effective area
exceeding the BATSE effective area by a factor of two or more. 

Roughly, compared to $RHESSI$, an instrument as in Table~1 would have 6 times more 
geometrical area, but the gain in effective area {\it for double events} should
be much larger. The background due to chance coincidence of single events would
be negligible, compared to $RHESSI$ which has a time window $\tau$=5 $\mu$s,
as in RF03 \footnote{The precise value of $\tau$ in CB03 is not given.}. 
For example, assuming $\mu \approx$ 20\%, a gain over $RHESSI$ of a factor of 10
in effective area, $B \approx S/2$ for $RHESSI$ (as in CB03, but RF03 suggest a 
much larger $B$), and $B \ll S$ for the proposed instrument, we would obtain
from Eq.~4
$$  
\left(\Pi _{min} \right)_{RHESSI} = n _{\sigma} \frac{\sqrt{3S}}{\mu S} = n
_{\sigma} \frac{\sqrt{3}}{\mu \sqrt{S}} \approx 4~ \left(\Pi _{min} \right)
_{this~work}  
$$
i.e. a simple instrument as the one sketched in this work would be more
sensitive to polarization than $RHESSI$, at a fraction of the cost.

\begin{figure}[!ht]
\epsfig{file=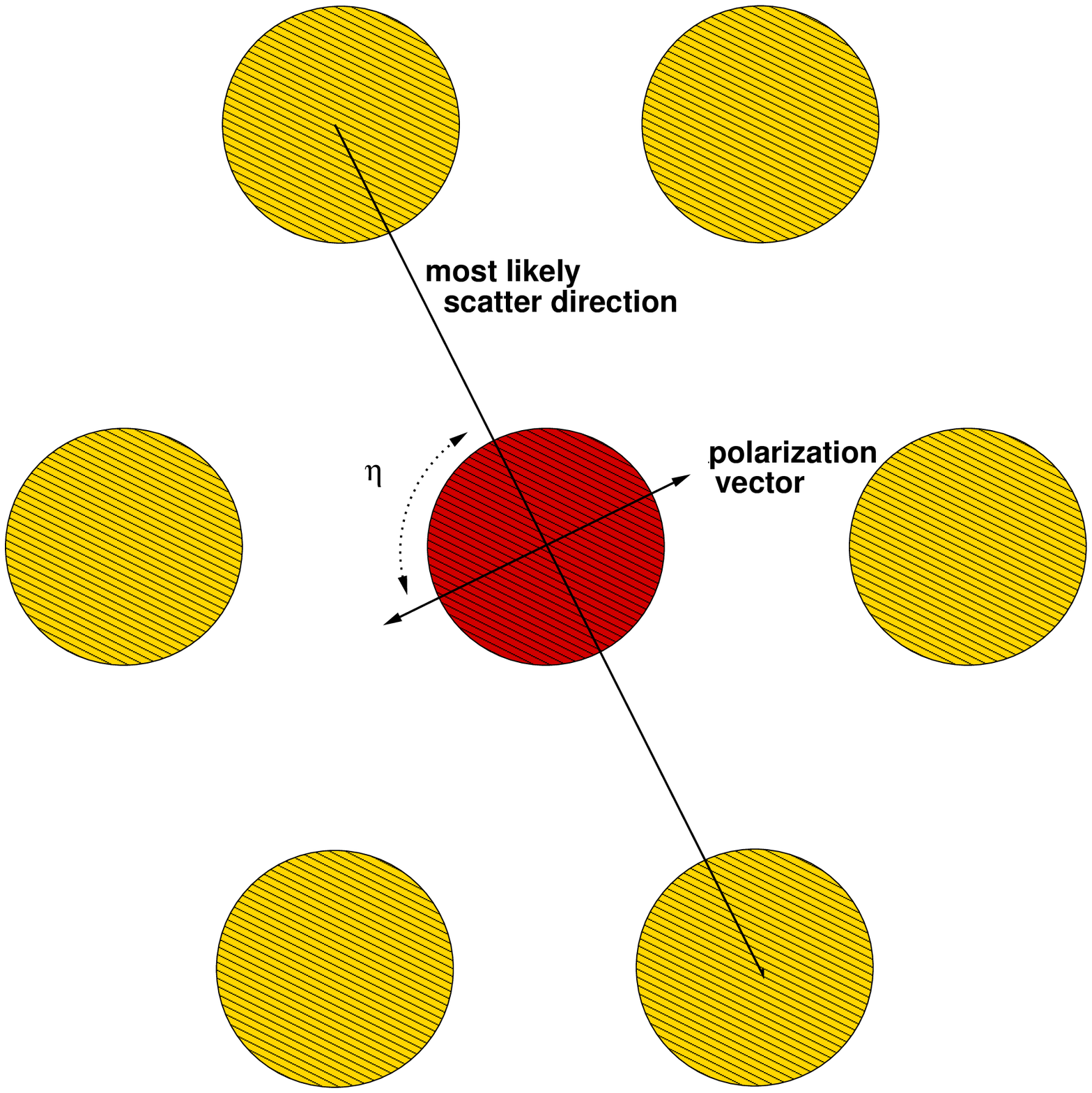,width=7cm}
\hfil
\epsfig{file=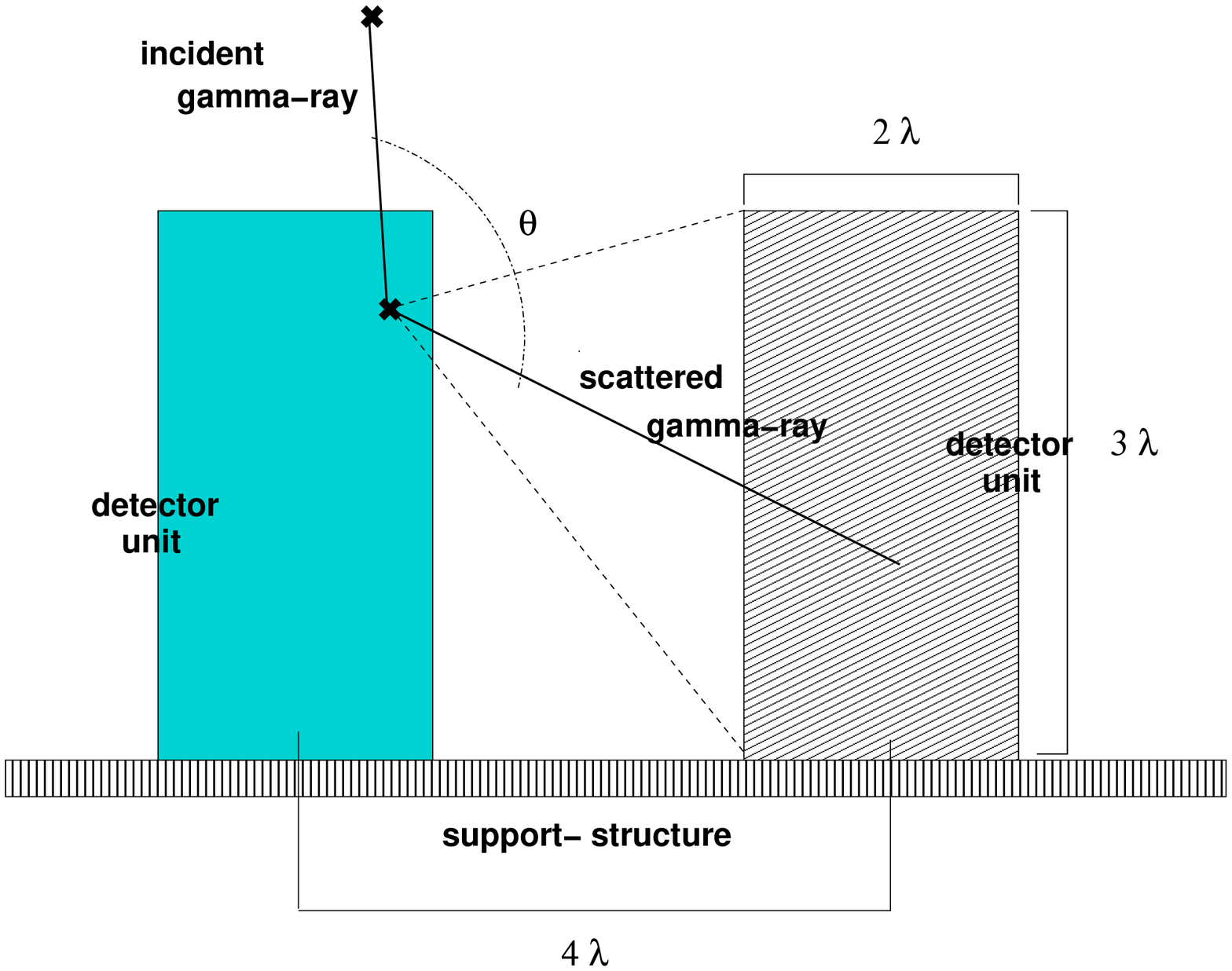,width=7cm}	
\caption{\label{fig1} Sketch of a $RHESSI$-like experimental apparatus for
polarimetry measurements. {\it Left:} top view. {\it Right:} side view of two
detector units; $\theta$ is the Compton scatter angle, $\lambda$ is ``mass
attenuation length''. }
\end{figure}

\begin{table}
\begin{center}
\footnotesize{
\begin{tabular}{|c|c|c|c|}\hline
instrument                  &  BATSE     &  RHESSI    & this work$^c$ \\ \hline \hline
effective area$^a$ [cm$^2$] & 2000       &  360       & 2000          \\ \hline
background$^b$ [Hz]         & --         &  500       & 1             \\ \hline \hline
\end{tabular}
}
\medskip
\medskip
\caption{\label{tab1} Comparison between $RHESSI$, BATSE and the present
work; all the figures are roughly accurate. $^a$: the effective area, calculated
from Eq.~6, is essentially the geometrical area. $^b$: the background is defined
as in Eq.~5, assuming an interaction rate of 10 kHz; for $RHESSI$, we assume
$\tau$=5 $\mu$s, as in RF03. $^c$: we assume $\tau$=10 ns and a geometrical area
of 2000 cm$^2$, as discussed in the text.}
\end{center}
\end{table}

\section{Feasibility of a Balloon Borne Experiment}

A balloon payload is an attractive option to carry GRB polarization measurements 
because of its minimal operation cost and complexity compared to a satellite
mission, and the possibility to come on line in a much shorter time. 
If a balloon borne experiment can achieve a field of view as large as BATSE 
\footnote{One should take care that the field of view is large not only for
triggering a GRB but also for measuring polarization, a rather demanding
requirement. For example, a cubic instrument with five sides similar to the
configuration proposed here may come close to solving the problem}, given a
comparable effective area, it would trigger as many GRBs as BATSE, i.e. about
one GRB per day based on the rate of events in the BATSE 4B catalog
\cite{paciesas}. 
Collecting a sample of at least $\sim$10 GRBs appears within reach for a 
long-duration balloon flight lasting for more than 20 days \footnote{Launched
from Antarctica, the HIREGS balloon borne experiment \cite{boggs} lasted more
than 20 days. Balloon flights over Antarctica are troublesome because of a high
background level in the MeV, but may be compatible with the loose background
requirement of the present work. } , or even through the combination of several
shorter balloon flights (e.g. \cite{lxe02} which lasted 27~h). Whether such a
sample would allow at least one significant polarization measurement in the
whole GRB sample is impossible to predict without the entailing analysis of a
final design.    
Clearly, the success of a balloon experiment will depend not only on the details
of the instrument but also in the brightness and hardness of the GRB
sample; GRB 021206 was an extraordinarily bright event, with a fluence of 1.6
10$^{-4}$ ergs cm$^{-2}$ \cite{hurley1} placing it in the upper tier of the
BATSE catalog \cite{paciesas}.  
Making the bold assumption that the signal $S$ scales as the fluence, the
proposed instrument should be able to return solid measurements for at least one
GRB event over 20 days.  
An estimate may be based on the peak flux of known GRBs in units of ph cm$^{-2}$ 
s$^{-1}$, as shown in Fig.~3 (from \cite{paciesas}). Eq.~4 may be turned around
to give $S$, fixing $\Pi _{min}$=50\%, $n _{\sigma}$=3 and $B \approx S$, with
$B$ mainly determined by the atmospheric \g-ray flux  
\begin{equation}
S = \left( \frac{2 n _{\sigma}}{\mu \Pi _{min}} \right)^2 = 3600
~\mathrm{double~events} 
\end{equation}
$S$ can be written in terms of the peak flux $PF$ as
\begin{equation}
S = T[\mathrm{s}] \times PF[\mathrm{ph}~\mathrm{cm}^{-2}~\mathrm{s}^{-1}]
\times f \times A_{eff}[\mathrm{cm}^{2}]
\end{equation}
where $T$ is a proper integration time and $f \times A_{eff}$ is the
effective area for double events (with $f < 1$ by definition). Assuming some 
reasonable values such as $T$=2~s, $f$=10\% and $A_{eff}$=2000~cm$^2$, a peak
flux of $\sim$10 ph cm$^{-2}$ s$^{-1}$ would allow to test $\Pi _{min} >$ 50\%
at the  3$\sigma$ level. From Fig.~3 it looks perfectly plausible that
polarization  should be detectable in about 10\% of the triggered GRBs.
\begin{figure}[!ht]
\centering
\epsfig{file=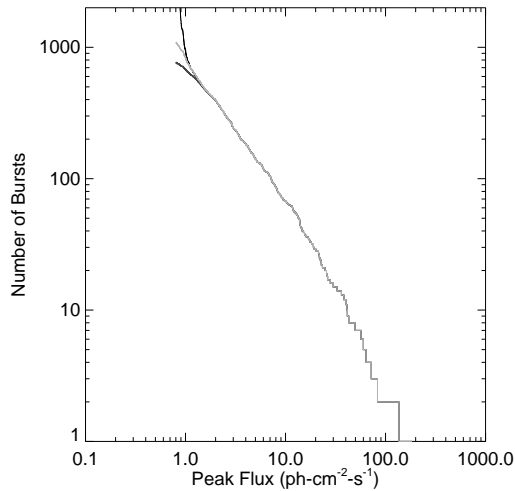,width=7cm}
\caption{\label{fig-flu} An example of log$N$-log$P$ integral distribution for
GRBs in the BATSE 4B catalog (from \cite{paciesas}). }
\end{figure}

An additional requirement for a successful polarimetry experiment rests on its
ability to determine the GRB localization. It is important to stress that, while
the source position is usually well known a priori in polarization measurements,
it is not the case for GRBs.
The GRB position in the sky should be known in order to define the azimuthal
distribution and modulation of Compton scattered photons. It also enters the
analysis in ruling out possible asymmetries that may mimic polarization. 
A detailed analysis indicates that a relatively coarse localization should
suffice. Such a localization can be achieved by the instrument itself, improving
over BATSE.  
Alternatively, a combination of already existing instruments (HETE, IPN) and the
upcoming SWIFT and GLAST missions may provide enough sensitivity for an
independent localization of the GRB.    
A successful polarization experiment must also be able to properly account for
systematic uncertainties such as asymmetries in the mass distribution of the
instrument. This can be accomplished either by careful Monte Carlo simulations,
or alternatively by rotation of the detector, as for $RHESSI$
\cite{mcconnell2}. 
Details on this notoriously delicate issue will require a careful analysis in
the future.  

An alternative to a balloon borne experiment is to exploit a future satellite
mission where a polarization instrument might ``piggyback'' another instrument,
possibly with imaging capability in soft \g-rays, which would significantly
increase the exposure and offer a good localization capability. 
Notwithstanding, a balloon phase is vital to provide proof of concept for an
instrument with an effective area comparable to BATSE.  

\section{Conclusions and Future Prospects}

The next generation of proposed instruments with the capability of measuring
polarization in soft \g-rays, including ACT \cite{kurfess,ACT:proposal:2003},
GRAPE \cite{mcconnell2}, and the design discussed here, promise important
constraints of the prompt GRB emission mechanism.    
While the ACT and GRAPE proposals have a much broader scope than just measuring
GRB polarization, the concept put forward in this work is the one of an
instrument which will just measure GRB polarization.
Even an upper limit on polarization will impose stringent limits for different
GRB models. Polarization measurements might also provide a novel way to explore
the mechanism responsible for the hard, short class of GRB events since the
instrumental resolution is not limited to long events. In particular, it is
important to explore if there is more than one class of GRBs in terms of
polarization.  
To assess the potential of a future experiment requires a full-fledged Monte
Carlo simulations including a detector mass model, GRB flux distribution and a
proper background model. This is of vital importance when evaluating the need
for rotation of the detector and modeling the sensitivity of large area
experiments. 
The estimates given in this work are based on simple experimental arguments,
scaling the performance of proven instruments like $RHESSI$ and BATSE, and
should be useful in setting the initial guidelines for new experimental
approaches to GRB polarization measurements. \\  
Summarizing, the minimal requirements in our analysis are: 
\begin{itemize}
\item[1. ] an effective area of 2000 cm$^2$ for ``triggering'' GRBs;
\item[2. ] enough sensitivity to a large fractional polarization for GRBs with
peak flux of $\sim$10 ph cm$^{-2}$ s$^{-1}$.
\end{itemize}

There is at least one last point that has been overlooked in most of the current
polarization literature including this work, which is the possibility of
energy measurements. Given a fully contained double event, Compton kinematics
must be obeyed and this constraint should improve background rejection.
In principle, it might even be possible to apply Compton imaging techniques
\cite{scho} to image the GRB, as would be the case (at a very refined level) for
the proposed ACT.     
Lastly, we have pointed out that the technology to take the next step in
polarimetry experiments is readily available and a working instrument may be
built in relatively short time. The simplicity of the proposed instrument
derives from the fact that its goal is narrowly focused on a  single,
well-defined measurement without stringent requirements on energy resolution,
source imaging capability and low background level. Alas, a simple instrument
does not imply a simple experiment.  

\section*{Acknowledgments}

We would like to thank Jules Halpern for reading the manuscript.

\bibliography{GRB-pol}  

\end{document}